\documentclass[letterpaper,10pt]{article} 

\usepackage{opticameet3} 

\newcommand\authormark[1]{\textsuperscript{#1}}

\usepackage{amsmath,amssymb}
\usepackage[colorlinks=true,bookmarks=false,citecolor=blue,urlcolor=blue]{hyperref} 
\usepackage{caption, subcaption}

\begin{document}

\title{Machine learning based joint polarization and phase compensation for CV-QKD}


\author{Hou-Man Chin,\authormark{1,2,*} Adnan A.E. Hajomer,\authormark{1} Nitin Jain,\authormark{1}, Ulrik L. Andersen,\authormark{1} and Tobias Gehring\authormark{1}}

\address{\authormark{1}Center for Macroscopic Quantum States, bigQ, Department of Physics, Technical University of Denmark, 2800 Kongens Lyngby, Denmark\\
\authormark{2}Machine Learning in Photonic Systems, Department of Photonics, Technical University of Denmark, 2800 Kongens Lyngby, Denmark}

\email{\authormark{*}homch@dtu.dk} 

\begin{abstract}
We investigated a machine learning method for joint estimation of polarization and phase for use in a Gaussian modulated CV-QKD system, over an 18 hour period measured on a installed fiber with 5.5 dB attenuation.
\end{abstract}

\section{Introduction}
Continuous-variable quantum key distribution (CV-QKD) is a field of research which has gotten increasing amounts of focus, for its guarantee of future proof information theoretic security but also due to its similarities with standard optical telecommunications systems. The milestone of realizing practical CV-QKD systems \cite{Jain2022Practical} is a important step towards commercialisation. One key aspect to maximize the secret key generation with a CV-QKD system is to align as perfectly as possible the polarization states of the quantum signal and the local local oscillator (LLO). This has been typically done through a polarization controller in laboratory conditions; either on a per measurement basis \cite{Jain2022Practical} or with automatic feedback \cite{Liu2020FPGAPC} with single polarization detection, more recently polarization diverse receivers and digital signal processing fulfill this purpose \cite{Pereira2021QCE}. To this end we propose to use a machine learning based framework for joint estimation and compensation of polarization state and phase noise with polarization diverse coherent receivers.

\section{Digital Signal Processing}
The impact of polarization rotation upon a CV-QKD system is reasonably straight forward and can be seen through the equation for secret key rate in the asymptotic regime \cite{Chin2021}

\begin{equation}
    s_{n} = \beta I(A:B) - \chi(B:E)\ ,
    \label{eqn:SKR}
\end{equation}
where $I(A:B)$ is the mutual information of the quantum signal transmitted from Alice (A) to Bob (B), $\beta$ is the reconcilication efficiency, and $\chi(B:E)$ is the Holevo information extracted by the attacker (Eve, E). The amount of Holevo information gained by Eve is determined for a given channel transmittance, by the \textit{excess noise} of the system. A misalignment of the quantum signal's polarization state with the LLO will reduce $I(A:B)$ and also increase $\chi(B:E)$, attributed to an increase in untrusted loss. In a practical system where phase noise is the dominant source of excess noise, this reduction in $I(A:B)$ will proportionally reduce excess noise. 

The channel model for an electric field of light transmitted through an optical fiber experiencing both polarization rotation and phase noise can be described using the following time varying equation \cite{Jignesh2016OE}
\begin{equation}
    \left[{\begin{array}{cc} E^{x}_{rx} \\
    E^{y}_{rx}\\ 
    \end{array}} \right] =  \left[{\begin{array}{cc}cos(\theta) & sin(\theta) \\
    -sin(\theta) & cos(\theta) \\
    \end{array}} \right]
    \left[{\begin{array}{cc} E^{x}_{tx} \\
    E^{y}_{tx}\\ 
    \end{array}} \right]exp(j\phi)\ ,
    \label{eqn:channelmodel}
\end{equation}
where $\theta$ is the angle of rotation and $\phi$ is the phase noise. Typically, these impairments would be compensated separately  via a constant modulus algorithm (CMA) for the polarization impairment and some method for phase compensation. We propose to use a joint polarization and phase estimation, machine learning based algorithm to compensate these impairments, specifically an unscented Kalman filter (UKF) implementation. Such a framework has been successfully used for phase compensation in CV-QKD \cite{Jain2022Practical}, demonstrating a performance advantage over standard methods \cite{Chin2021}, particularly in low signal-to-noise ratio regimes. Given the joint estimation, the process is therefore nonlinear due to the $\exp(j\phi)$ term, necessitating the UKF or similar Kalman filter variants. Such a method has also been previously demonstrated for classical optical telecommunications systems \cite{Jignesh2016OE}. The measurement equation we use for the UKF is as follows for given noisy measurements $y^{1,2}_{k}$ of both polarizations

\begin{equation}
    \left[{\begin{array}{cc} y^{1}_{k} \\
    y^{2}_{k}\\ 
    \end{array}} \right] =  \left[{\begin{array}{cc}a_{k} & b_{k} \\
    -b_{k} & a_{k} \\
    \end{array}} \right]
    \left[{\begin{array}{cc} \sqrt{P_{sig}} \\
    0\\ 
    \end{array}} \right]\cos(2\pi f + \phi_{k})\\ 
    +
     \left[{\begin{array}{cc} R^{1}_{k} \\
    R^{2}_{k}\\ 
    \end{array}} \right]
\end{equation}
where \textit{a} and \textit{b} are estimated variables for polarization rotation, $P_{sig}$ is the power of the estimated signal, $\phi$ is the estimated phase noise, \textit{R} is the measurement noise, \textit{f} is the frequency of the estimated signal, \textit{k} represents the sample estimated. Due to inherently low power of the quantum signal, a pilot tone is frequency multiplexed with it to enable digital signal processing. The state equation assumes that the current state is equivalent to the previous.
A one tap CMA will be used as the reference algorithm for comparison with the joint machine learning method, for phase compensation we will use with the CMA the phase noise only version of the UKF \cite{Chin2021}. For both algorithms, the rotation matrix is normalized such that the power before and after compensation remains the same.

\section{Experimental Setup}
\begin{figure}[tb]
    \centering
    \includegraphics[scale=1.05]{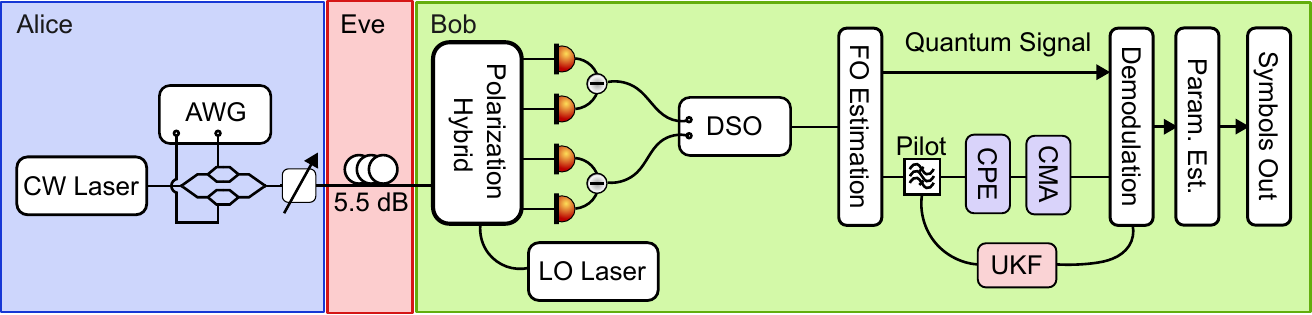}
    \caption{Experimental setup, AWG - arbitrary waveform generator, DSO - digital storage oscilloscope, CW - continuous wave, FO - frequency offset, CPE - carrier phase equalization, CMA - constant modulus algorithm, UKF - unscented Kalman filter, Param. Est. - parameter estimation}
    \label{fig:expsetup}
\end{figure}
\begin{figure}[tb]
     \centering
     \begin{subfigure}[b]{0.45\textwidth}
         \centering
         \includegraphics[width=\textwidth]{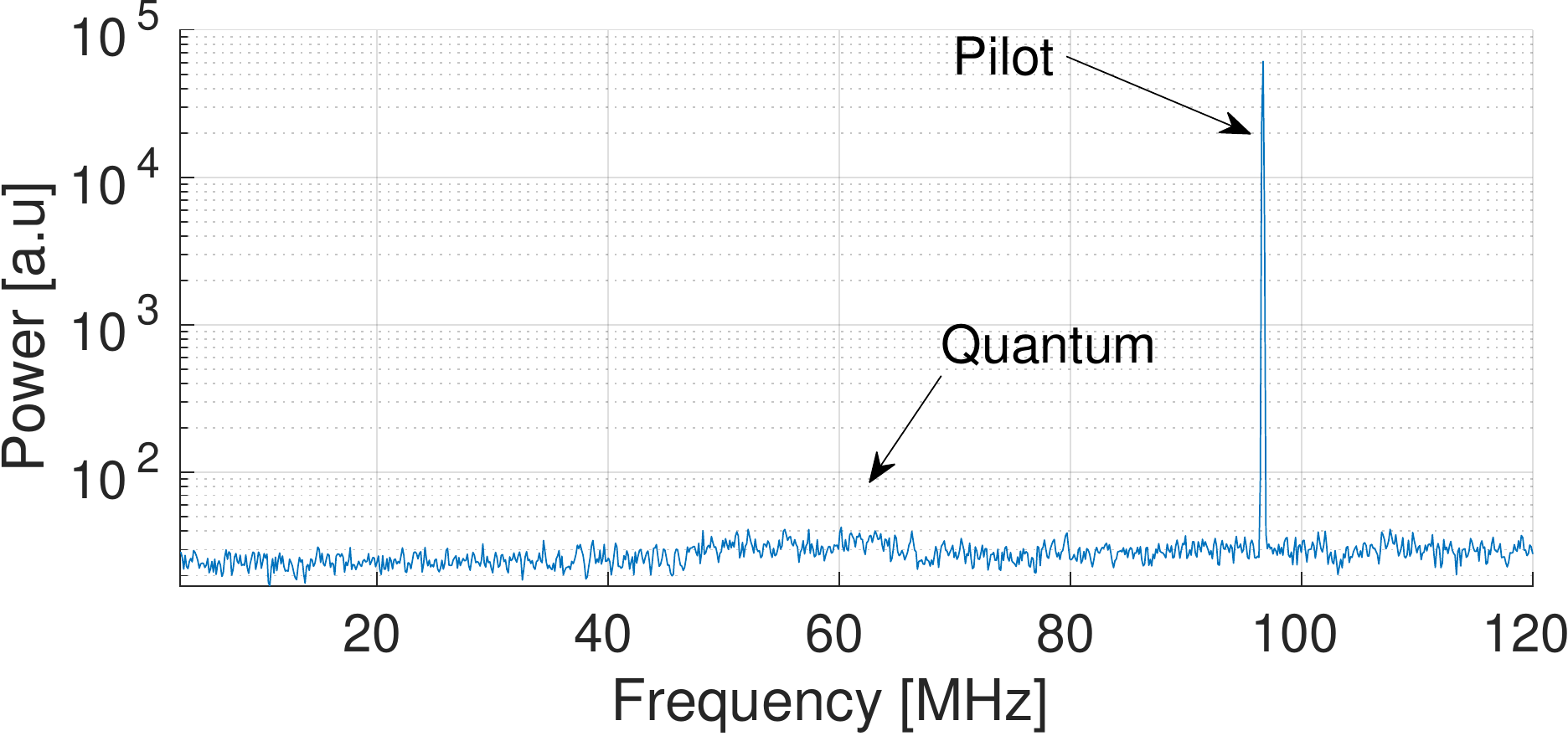}
         \caption{}
         \label{fig:RxPSD}
     \end{subfigure}
     ~
     ~
     \begin{subfigure}[b]{0.45\textwidth}
         \centering
         \includegraphics[width=\textwidth]{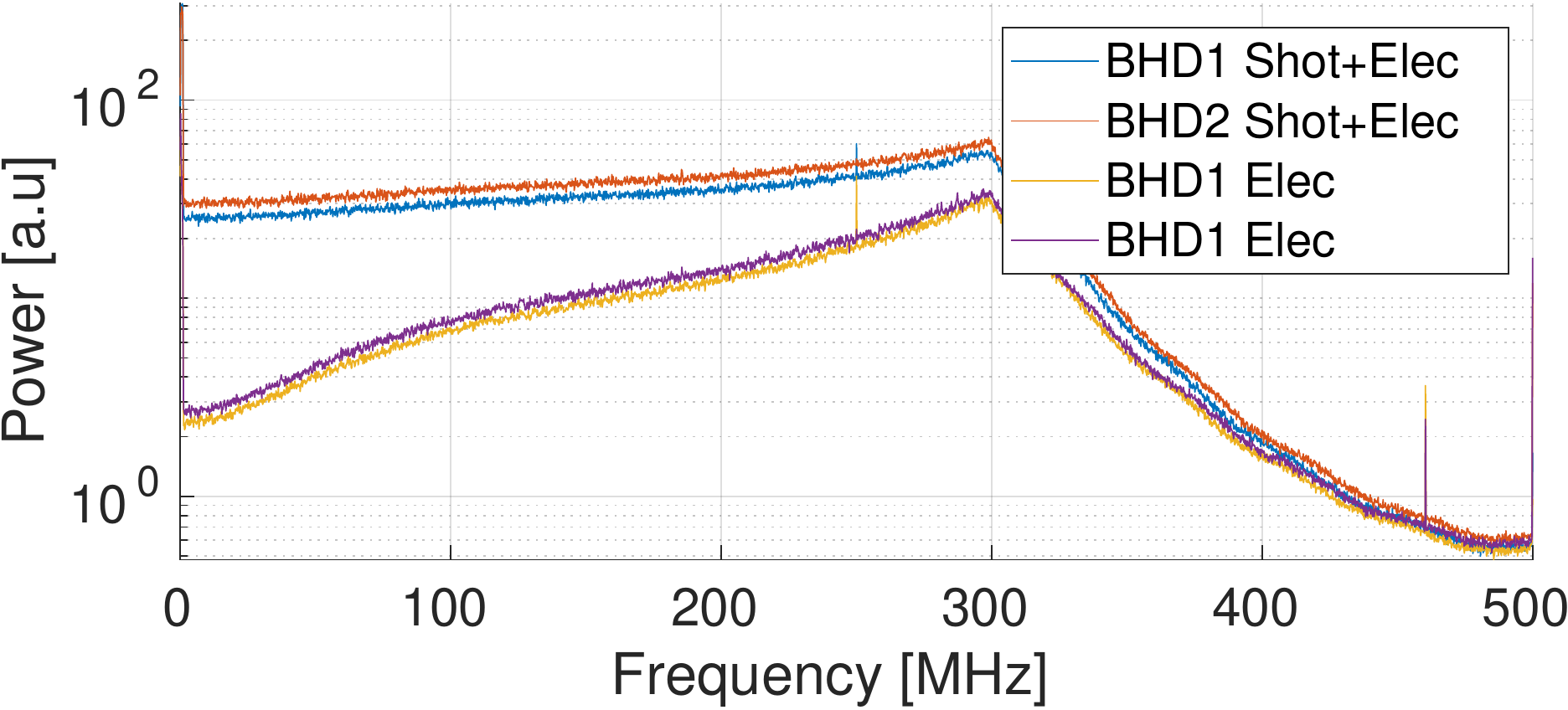}
         \caption{}
         \label{fig:RxDets}
     \end{subfigure}
        \caption{(a) Example spectra of the received quantum signal and pilot tone, (b) Spectral profiles of the balanced heterodyne detectors (BHD) for both polarizations}
        \label{fig:results}
\end{figure}
The experimental setup is as shown in Fig.~\ref{fig:expsetup}. We prepare quantum states of light using bits generated by a quantum random generator. The quantum signal is generated at 20 Mbaud, upsampled with a root raised cosine filter (roll-off $\sigma=0.2$) and frequency multiplexed with a pilot tone as in \cite{Chin2021}. A 1 GSample/s arbitrary waveform generator (AWG) modulates the combined waveform onto the output of a continuous-wave (CW) laser (100 Hz) with an IQ modulator. This optical signal was transmitted through an optical link running on campus with loss equivalent to 5.5 dB, see Fig.~\ref{fig:RxPSD} for an example of the received optical spectra. A polarization hybrid separates orthogonal polarizations of the received signal which were detected with balanced RF heterodyne detection. The receiver has overall trusted loss $\tau \approx0.53$ and respective responses can be seen in Fig~\ref{fig:RxDets}. The local oscillator (LO) is generated by another 100 Hz CW laser. Electronic and shot noise calibrations were performed with all lasers off and only LO respectively. The transmitted signal modulation variance ($N=1.65$ SNU) is calibrated in a back to back configuration. A set of measurements were performed hourly consisting of $ 4.9 \times 10^{5}$ symbols in frames of 10k symbols.

\section{Results}
\begin{figure}[htb]
     \centering
     \begin{subfigure}[b]{0.38\textwidth}
         \centering
         \includegraphics[width=\textwidth]{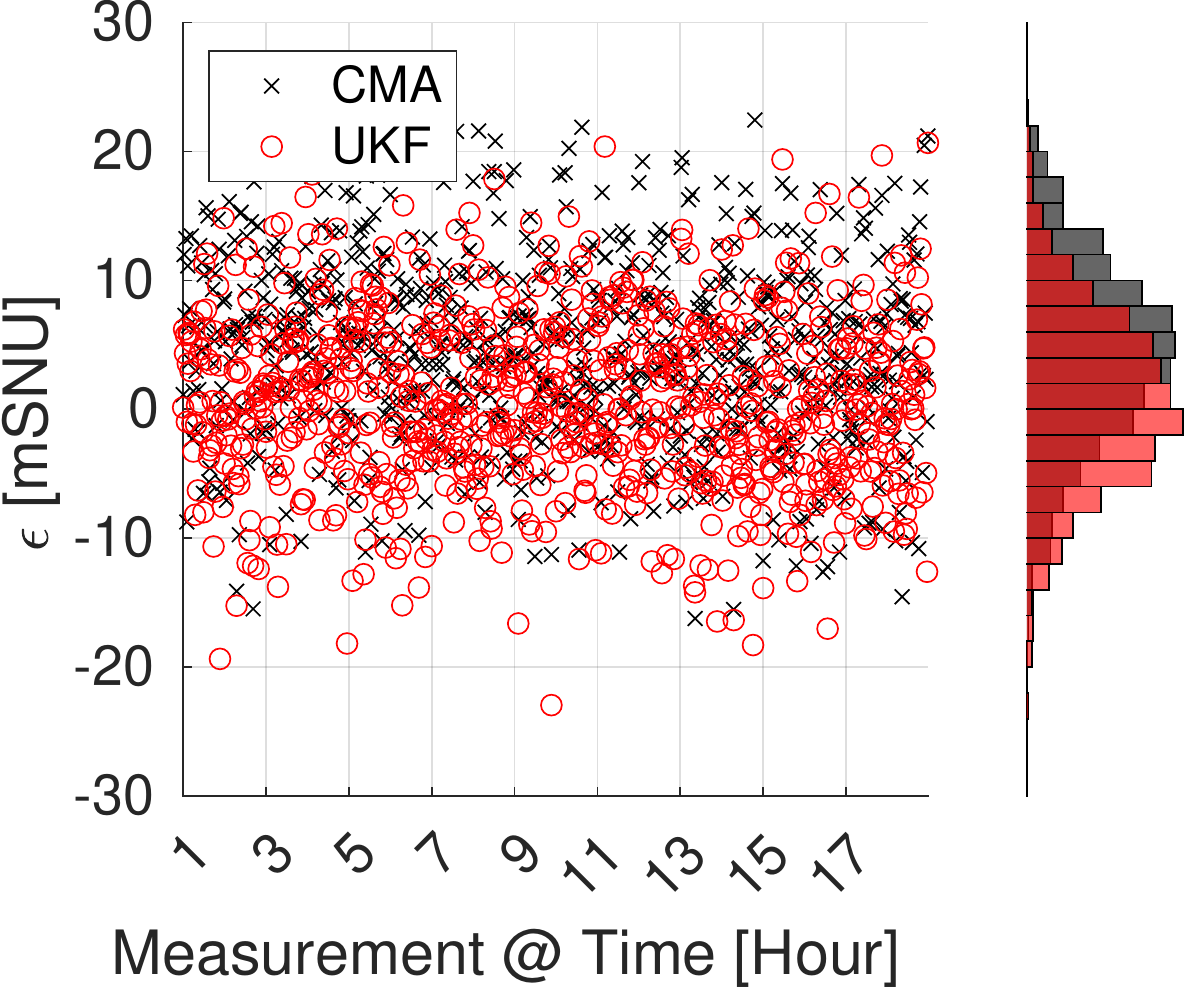}
         \caption{}
         \label{fig:EN}
     \end{subfigure}
     \hfill
     \begin{subfigure}[b]{0.6\textwidth}
         \centering
         \includegraphics[width=\textwidth]{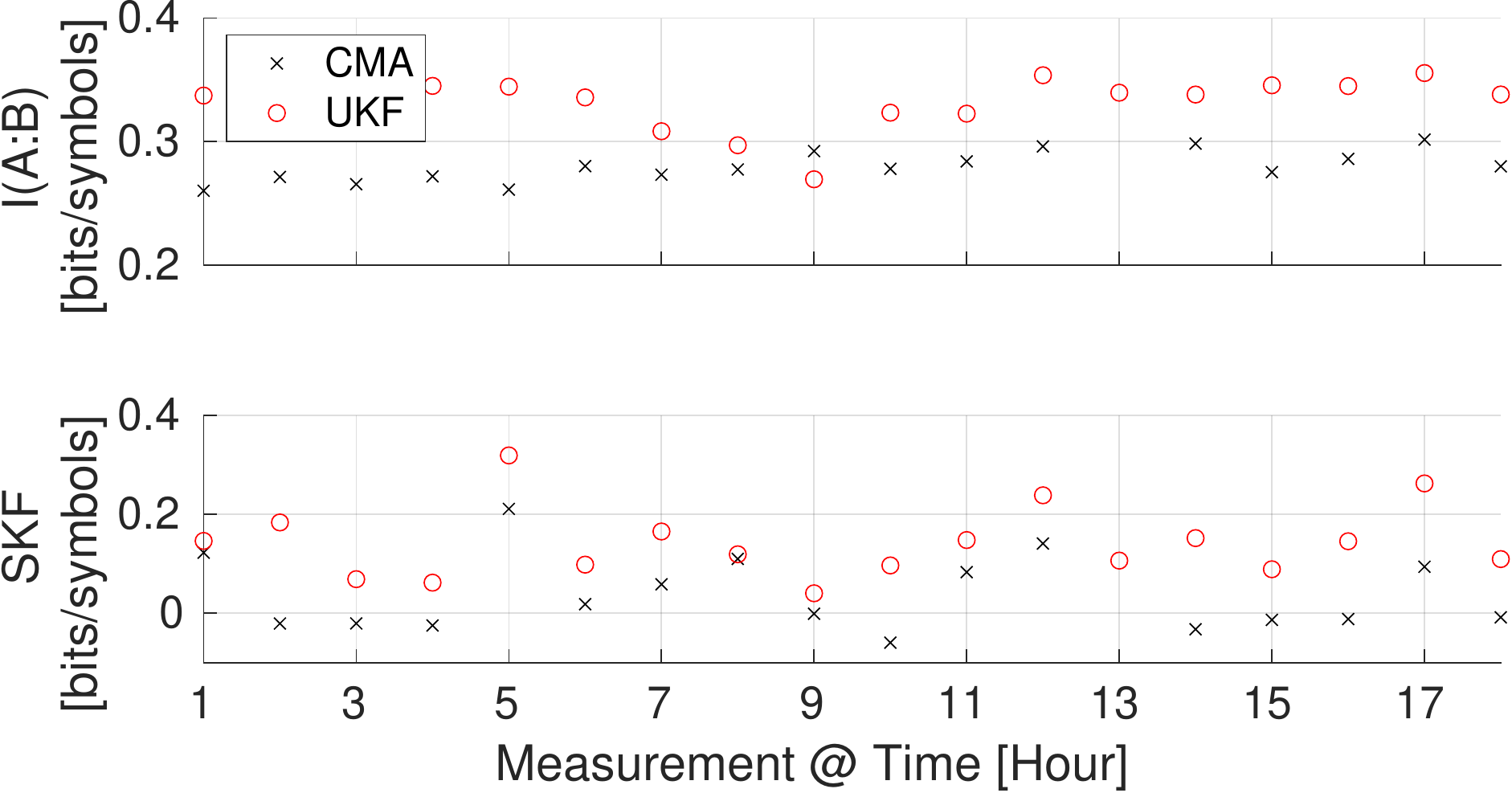}
         \caption{}
         \label{fig:MI_SKF}
     \end{subfigure}
        \caption{Results for each set of hourly measurements when using both methods, (a) excess noise of the same symbols split into frames of 10k symbols, (b, top) the mutual information of Alice to Bob $I(A:B)$ averaged for the 490k symbols per measurement, (b, bottom) asymptotic secret key fraction (SKF) achieved for these measurements, assuming 95\% reconciliation efficiency}
        \label{fig:results}
\end{figure}

The received signals were equalized according to their respective detector responses and also normalized with respect to each other for power over 300 MHz. The CMA convergence parameter was $\mu=0.01$, other values for $\mu$ traded off mutual information with excess noise. Each set of measurements performed hourly were processed on a frame level of 10k quantum symbols. Fig~\ref{fig:MI_SKF} (top) shows the mutual information achieved by both the UKF and reference CMA. It is clear that the UKF for joint polarization and phase compensation out performs the CMA and phase compensation UKF. We attribute the disparency in mutual information due to the CMA converging slower than the UKF. The excess noise is shown in Fig.~\ref{fig:EN} for each measured frame, the corresponding x-axis merely denotes at which measurement time the frame was taken. The mean  excess noise is $\approx 4.9$ mSNU with the CMA and $\approx 0.6$ mSNU with UKF, this is reflected in the histograms of per frame excess noise shown on the right of Fig.~\ref{fig:EN}. We believe this is due to imperfection in normalization with the CMA's estimated coefficients with respect to the UKF. With a longer frame length, a similar or smaller $\mu$ might allow the CMA to perform better. Fig.~\ref{fig:MI_SKF} (bottom) shows the secret key fraction per transmitted symbol. Using joint estimation of the polarization state and phase noise yielded a consistently positive secret key fraction while with the CMA, positive key was achieved only 9 measurement sets out of 18. This secret key fraction was calculated under the assumption of asymptotic regime and 95\% reconciliation efficiency. We note that the modulation variance was not optimized for this particular link.

\section{Conclusion}
We examined the performance of a machine learning based framework implementing an unscented Kalman filter for joint estimation of polarization rotation and laser phase noise. Its performance was compared to a standard CMA combined with an unscented Kalman filter for phase estimation for operation of a CV-QKD system with Gaussian modulated coherent states. Significantly greater mutual information and secret key fraction was achieved, yielding a consistently positive rate of secret key generation compared to our reference method which failed to produce a positive key in 9 out of 18 hourly measurement sets. We believe that using such a machine learning framework is an option for systems seeking to generate the maximum possible secret key at the cost of computational complexity.

\section*{Acknowledgements}
The authors acknowledge financial support from Innovation Fund Denmark (CryptQ project  \#0175-00018A), OPEN-QKD (\#857156) and the DNRF, Center for Macroscopic Quantum States (bigQ, DNRF142), and DCC \cite{DTU_DCC_resource}.


\begin{thebibliography}{1}
\footnotesize
\newcommand{\enquote}[1]{``#1''}

\bibitem{Jain2022Practical}
N.~Jain et. al., \enquote{{Practical continuous-variable quantum key distribution
  with composable security},} {\protect\JournalTitle{Nature Communications}}
  \textbf{13} (2022).

\bibitem{Liu2020FPGAPC}
W.~Liu et. al., \enquote{{Continuous-variable quantum key
  distribution under strong channel polarization disturbance},}
  {\protect\JournalTitle{Physical Review A}} \textbf{102}, 32625 (2020).

\bibitem{Pereira2021QCE}
D.~Pereira et. al., \enquote{{A polarization diversity
  CV-QKD detection scheme for channels with strong polarization drift},} in
  \emph{IEEE Intern. Conf. on Quantum Computing and Engineering,}  vol. 102
  (IEEE, 2021), pp. 469--470.

\bibitem{Chin2021}
H.-M. Chin et. al., \enquote{Machine
  learning aided carrier recovery in continuous-variable quantum key
  distribution,} {\protect\JournalTitle{npj Quantum Information}} \textbf{7}
  (2021).

\bibitem{Jignesh2016OE}
J.~Jignesh et. al., \enquote{{Unscented Kalman
  filters for polarization state tracking and phase noise mitigation},}
  {\protect\JournalTitle{Optics Express}} \textbf{24}, 1597--1607 (2016).

\bibitem{DTU_DCC_resource}
{DTU Computing Center}, \enquote{{DTU Computing Center resources},}  (2021).

\end{thebibliography}
\end{document}